\documentclass[a4paper]{article}
\usepackage[psamsfonts]{amssymb}
\usepackage{amsmath}
\usepackage{epsfig}

\author{H. Mohseni Sadjadi\footnote{mohseni@phymail.ut.ac.ir}
\\ {\small Department of Physics, University of Tehran ,}
\\ {\small P. O. B. 14395-547, Tehran 14399-55961, Iran}}
\title{ Multicomponent solution in modified theory of gravity in the early universe}

\begin{document}
\maketitle
\begin{abstract}
We study the modified theory of gravity in Friedmann Robertson
Walker universe composed of several perfect fluids. We consider
the power law inflation and determine the equation of state
parameters in terms of the parameters of modified gravity's
Lagrangian in the early universe. We also discuss briefly the
gravitational baryogenesis in this model.
\end{abstract}
\section{Introduction}
Modified theory of gravity, constructed by adding geometrical
correction terms to the usual Einstein Hilbert Lagrangian has been
used to study the inflationary epoch and the subsequent reheating
stage in the early universe \cite{stra,Mij}. In this view the
{\it{effective Lagrangian}}, $L$, in the early universe includes
higher order curvature terms and inflation may be a natural result
of this theory. In \cite{stra}, it was shown that involving a term
proportional to the square of the scalar curvature, i.e.,
$L=R+\alpha R^2$, results in a quasi-de Sitter expansion. In this
model the Hubble parameter decreases slowly for large $\alpha$
before going into an oscillation phase which can reheat the
universe \cite{Mij}. In principle, one can assume that the
effective Lagrangian is a function of the scalar curvature
$L=f(R)$ \cite{barrow}. The time dependence of the scale factor
depends on $f(R)$, e.g., if one chooses $f(R)=R+\alpha R^2+\beta
R^3$, instead of an exponential like inflation, he may obtain a
power law expansion for the universe \cite{Berk}. Some of particle
physics problems such as the hierarchy problem \cite{noj} and
baryogenesis \cite{Lamb} can also be studied in the framework of
modified theory of gravity. Recently, the modified theory of
gravity which is able to describe the present cosmic acceleration
\cite{acc} without involving exotic dark energy \cite{mod} has
attracted more attention.

In this paper we aim to study the universe in an era when the
expansion can be described by the scale factor $a(t)\propto
t^\lambda$.  This assumption is consistent with most parts of the
(Friedmann Robertson Walker expanding) universe history, when one
of the components such as (dark) matter or radiation dominates.
But in the framework of the usual theory of gravity, and in the
presence of ordinary matter and radiation, the power law expansion
can not describe the inflationary and accelerating expansion
phases. Besides, the gravitational baryogenesis proposal fails to
determine the baryon asymmetry in radiation dominated era. In this
paper we consider the modified theory of gravity, characterized by
$f(R)=\sum_iA_iR^{n_i}$, in the early universe. We assume that the
universe is composed of different perfect fluid components with
constant equation of state parameters. Physical parameters of the
system such as radiation temperature and the equation of state
parameters of the fluids are determined in terms of $n_i$'s and
$A_i$'s. At the end we use our results to determine the
gravitational baryogenesis in a universe composed of two
components in the framework of modified theory of gravity. Through
the paper we use the units $c=k_B=1$.

\section{Power law expansion in  modified theory of gravity}
We consider the modified theory of gravity described by the action
\begin{equation}\label{1}
S=\int \big({1\over {16\pi G}}f(R)+L_m\big)\sqrt{-g}d^4x,
\end{equation}
where $L_m$ is the Lagrangian corresponding to the matter such as
radiation, baryonic matter, dark matter and so on. The geometry of
the universe which is assumed to be spatially flat, homogeneous
and isotropic is described  by Friedmann Robertson Walker (FRW)
metric
\begin{equation}\label{2}
ds^2=-dt^2+a^2(t)(dx^2+dy^2+dz^2).
\end{equation}
$a(t)$ is the scale factor in terms of which the Hubble parameter
is $H={\dot{a(t)}\over a(t)}$. The Ricci scalar is
\begin{equation}\label{3}
R=6(\dot{H}+2H^2).
\end{equation}
By variation of the action with respect to the metric, we obtain
\begin{equation}\label{4}
g_{\mu \nu}\Box f'(R)-\nabla_\mu \nabla_\nu f'(R)+f'R_{\mu
\nu}-{f(R)\over 2}g_{\mu \nu}=8\pi GT_{\mu \nu}^m,
\end{equation}
where $T^m_{\mu \nu}$ are the energy momentum tensor components of
matter fields which behave like a perfect fluid. The energy
density, $\rho$, and the pressure, $P$, may be derived from
(\ref{4}):
\begin{eqnarray}\label{5}
8\pi G\rho&=&{f(R)\over 2}+3{\dot{a}\over
a}f''(R)\dot{R}-3{\ddot{a}\over a}f'(R)\nonumber \\
8\pi GP&=&-{f(R)\over 2}+(2{\dot{a}^2\over a^2}+{\ddot{a}\over
a})f'(R)-2{\dot{a} \over
a}f''(R)\dot{R}-f'''(R)\dot{R}^2-\nonumber
\\
&&f''(R)\ddot{R}.
\end{eqnarray}
The energy conservation relation,
\begin{equation}\label{6}
\dot{\rho}+3H(P+\rho)=0,
\end{equation}
is not independent of (\ref{5}), but is required for consistency.
We assume that the perfect fluid is effectively composed of
non-interacting components, although each component may include
subcomponents whose interactions are rapid compared to the
expansion rate keeping them in thermal equilibrium. In this way
the total energy density and the total pressure are given by
\begin{equation}\label{7}
\rho=\sum_i \rho_i,\,\,\,P=\sum_iP_i.
\end{equation}
Each component satisfies
\begin{equation}\label{8}
\dot{\rho_i}+3H(P_i+\rho_i)=0.
\end{equation}
For time independent equation of state parameters (EOS) (denoted
by $\gamma_i$'s), the solution of (\ref{8}), in terms of the scale
factor, is
\begin{equation}\label{9}
\rho_i(t)=\rho_i(t_0)\Big({a(t)\over
a(t_0)}\Big)^{-3(\gamma_i+1)}.
\end{equation}
Note that even for constant $\gamma_i$'s, the EOS parameter of the
universe,
\begin{equation}\label{10}
\gamma={P\over \rho}={\sum_i\gamma_i\rho_i\over \sum_i\rho_i},
\end{equation}
is time dependent. To study (\ref{5}), let us take
\begin{equation}\label{11}
f(R)=\sum_i  f_i(R),
\end{equation}
such that $f_i$ satisfies
\begin{eqnarray}\label{12}
8\pi G\rho_i&=&{f_i(R)\over 2}+3{\dot{a}\over
a}f_i''(R)\dot{R}-3{\ddot{a}\over a}f_i'(R)\nonumber \\
8\pi GP_i&=&-{f_i(R)\over 2}+(2{\dot{a}^2\over a^2}+{\ddot{a}\over
a})f_i'(R)-2{\dot{a} \over
a}f_i''(R)\dot{R}-f_i'''(R)\dot{R}^2\nonumber \\
-f_i''(R)\ddot{R}.
\end{eqnarray}
In this way if (\ref{12}) is satisfied, then (\ref{5}) will be
also satisfied. In this paper we restrict ourselves to the models
characterized by \cite{odint}
\begin{equation}\label{13}
f(R)=\sum_iA_iR^{n_i}.
\end{equation}
$A_i$'s and $n_i$'s are real constants ($n_i$'s are not restricted
to integer numbers). Note that $f(R)$ broken power law models such
as \cite{saw}
\begin{equation}\nonumber
f(R)=R-m^2\frac{c_1({R\over m^2})^n}{c_2({R\over m^2})^n+1},
\end{equation}
where $m,n>0,c_1,c_2$ are real numbers, can be casted to
(\ref{13}) at high curvature limit $R\gg m^2$.

$f(R)$ in (\ref{1}) must satisfy some stability conditions
\cite{stab}. For the model (\ref{1}), the condition ${d^2f(R)\over
dR^2}>0$, which in our model reduces to
\begin{equation}\label{60}
\sum_in_i(n_i-1)A_iR^{n_i-2}>0,
\end{equation}
is necessary for classical stability of FRW solution in
high-curvature regime. In the context of quantum mechanics this
condition ensures the absence of tachyonic scalarons \cite{stra}.
We also require that ${df(R)\over dR}>0$ or
\begin{equation}\label{61}
\sum_in_iA_iR^{n_i-1}>0,
\end{equation}
to prevent the graviton from turning into a ghost \cite{stab}.

In the following We use the ansatz $a(t)\propto t^\lambda$ for the
scale factor. This ansatz is allowed when the number of (dominant)
fluid components is the same as the number of the terms in $f(R)$.
In general, $\lambda$ depends on EOS parameters of the components
of the perfect fluid, $\gamma_i$'s, and $n_i$'s. Now let us find
the conditions required for consistency of $a(t)\propto t^\lambda$
with (\ref{12}). By substituting the Ricci scalar,
$R=6\lambda{2\lambda-1\over t^2}$, and (\ref{9}) into (\ref{12})
we obtain
\begin{equation}\label{14}
(2\lambda-1)^{n_i-1}(6\lambda)^{n_i}\Big(n_i(3-\lambda)-2n_i^2-1+2\lambda\Big)
A_it^{-2n_i}= 16\pi G\rho_i(t_0)\Big({t\over t_0}
\Big)^{-3\lambda(\gamma_i+1)}.
\end{equation}
The above equation is true for all $t$, provided that
\begin{equation}\label{15}
{3\lambda\over 2}={n_i\over 1+\gamma_i}={n_j\over 1+\gamma_j},
\,\,\, \forall i,j,
\end{equation}
and
\begin{equation}\label{16}
A_i={\rho_i(t_0)t_0^{3\lambda(\gamma_i+1)}\over \delta_i},
\end{equation}
where
\begin{equation}\label{17}
\delta_i= {3\over 8\pi
G}\left(n_i(-\lambda+3)-2n_i^2+2\lambda-1\right)\left(6(2\lambda-1)\right)^{n_i-1}
\lambda^{n_i}.
\end{equation}
Hence if the EOS parameters of the fluid components satisfy
(\ref{15}), and $A_i$'s are given by (\ref{16}), then in $f(R)$
gravity specified by (\ref{13}), $a(t)\propto t^{\lambda}$ can be
considered as the scale factor. If $n_l=0$, then we must take
$\gamma_l=-1$ describing a cosmological constant corresponding to
the vacuum energy. For positive $\lambda$, $n_i<0$ leads to
$\gamma_i<-1$ describing a phantom like dark energy component.

The above discussion can be generalized to the case that $\lambda$
is a slowly varying function of time, $\lambda=\lambda(t)$,
$\dot{\lambda(t)}t\ll \lambda(t)$ \cite{odint}.  Time independence
of $A_i$ requires
\begin{equation}\label{18}
{\dot{\rho_i}\over \rho_i}={(n_i-2)\dot{R}\over R}+{\dot{Q_i}\over
Q_i},
\end{equation}
where $Q_i$ is defined through
$Q_i=R^2-6n_i\left(\dot{H}+H^2\right)R+6n_i(n_i-1)H\dot{R}$. By
$\gamma_i=-1-{\dot{\rho_i}\over 3H\rho_i}$  and (\ref{18}), we can
determine the time dependent EOS parameters
\begin{eqnarray}\label{19}
\gamma_i&=&-{3\lambda-2n_i\over
3\lambda}+{4(4\lambda-1)n_i^2+(4\lambda^2-17\lambda+2)n_i+11\lambda-8\lambda^2\over
3\lambda(2\lambda-1)(2n_i^2+(\lambda-3)n_i+(1-2\lambda))}{\dot{\lambda}t\over
\lambda}\nonumber\\
&&+ \mathcal{O}\left({\dot{\lambda}^2t^2\over \lambda^2}\right).
\end{eqnarray}
Up to $\mathcal{O}\left({\dot{\lambda}t\over \lambda}\right)$, the
relation between $\gamma_i$'s may be obtained in a compact form
\begin{eqnarray}\label{20}
\gamma_i(t)&=&{n_i\over n_j}\gamma_j(t)+({n_i\over
n_j}-1),\,\,\,\forall i,j \nonumber \\
&=&-1+{2n_i\over 3\lambda(t)}.
\end{eqnarray}
It is worth to note that in a one component fluid, with EOS
parameter $\gamma$, we have $\gamma={2n\over 3\lambda}-1$, hence
if the universe is approximately filled with radiation (i.e
$\gamma={1\over 3}$) then $\lambda={n\over 2}$ and
$R=3n\left({n-1\over t^2}\right)$. Therefore for $n\neq 1$ we have
$R\neq 0$ and $\dot{R}\neq 0$ while in Einstein theory of gravity
we obtain $\dot{R}=0$. This is the note used in \cite{Lamb} to
show that in modified gravity the gravitational baryogenesis may
occur even in radiation dominated epoch.

Now assume that one of the thermal fluid components (e.g. the
radiation component) has temperature $T$, i.e., the subcomponents
of this component have nearly common temperature $T$. We assume
also that the density of each component is given by
\begin{equation}\label{21}
\rho_i=\epsilon_iT^{\theta_i}.
\end{equation}
This is allowed when $T$ is proportional to a power of the scale
factor.  It is worth to note that $T$ may not be the temperature
of other noninteracting fluid components. (\ref{9}) and (\ref{16})
can be used to obtain the time dependence of $T$
\begin{equation}\label{22}
T=\left(\delta_i{A_i\over \epsilon_i }\right)^{1\over \theta_i}
t^{-2n_i\over \theta_i}.
\end{equation}
This equation holds for each $i$, therefore
\begin{equation}\label{23}
 {n_i\over \theta_i}={n_j\over \theta_j}\,\,\,\,\forall{i,j}.
\end{equation}
Besides, for $\forall i,j$ we must have
\begin{equation}\label{24}
\left({\delta_i\over \epsilon_i}A_i\right)^{1\over
\theta_i}=\left({\delta_j\over \epsilon_j}A_j\right)^{1\over
\theta_j}.
\end{equation}
To elucidate our results, as an example, we assume that the
universe is approximately composed of a radiation component
(denoted by the subscript $\mathcal{R}$) and a non thermal
component with EOS parameter $\omega$ \cite{david}. These
non-interacting components satisfy the energy conservation
relation
\begin{eqnarray}\label{25}
\dot{\rho_\mathcal{R}}+4H\rho_{\mathcal{R}}=0\nonumber \\
\dot{\rho_\omega}+3H(1+\omega)\rho_\omega=0.
\end{eqnarray}
The time derivative of the ratio of these components, denoted by
$r:={\rho_\mathcal{R}\over \rho_\omega}$, is
\begin{equation}\label{26}
\dot{r}=3Hr\left(\omega-{1\over 3}\right)
\end{equation}
if $\omega>{1\over 3}$, $\dot{r}>0$ and $\rho_\omega$ component
decreases more rapidly than radiation component and the universe
will become radiation dominated. Following our previous
discussions let us take
\begin{equation}\label{27}
f(R)=A_\mathcal{R}R^{n_\mathcal{R}}+A_{\omega} R^{n_\omega}.
\end{equation}
Stability conditions require that the parameters of the model
satisfy
\begin{equation}
n_\mathcal{R}(n_\mathcal{R}-1)A_\mathcal{R}R^{n_\mathcal{R}}+n_\omega(n_\omega-1)A_\omega
R^{n_\omega}>0,
\end{equation}
and
\begin{equation}\label{62}
n_\mathcal{R}A_\mathcal{R}R^{n_\mathcal{R}-1}+n_\omega A_\omega
R^{n_\omega-1}>0.
\end{equation}
If the scale factor is given by $a(t)\propto t^\lambda$, then
\begin{equation}\label{28}
n_\mathcal{R}=2\lambda ,\,\,\, n_{\omega}={3\over
2}(1+\omega)\lambda.
\end{equation}
As a result, in this model, $\omega$ can be expressed in terms of
$n_i$'s: $\omega={4n_\omega\over 3n_\mathcal{R}}-1$. The stability
conditions is satisfied by choosing appropriate parameters for the
model, e.g., if we take $n_\omega=1$, which results in
$\omega={2\over 3\lambda}-1$, (\ref{27}) reduces to $f(R)=A_\omega
R+A_\mathcal{R}R^{n_\mathcal{R}}$ (the stability of models
including this specific case was discussed in \cite{ojiri}) and
the stabilities conditions become
\begin{eqnarray}\label{63}
&&n_\mathcal{R}(n_\mathcal{R}-1)A_\mathcal{R}R^{n_\mathcal{R}}>0\nonumber
\\
&&n_\mathcal{R}A_\mathcal{R}R^{n_\mathcal{R}-1}+A_\omega>0.
\end{eqnarray}
For positive curvature and for $\lambda$'s belonging to the domain
$0.5<\lambda<0.64$, $A_\mathcal{R}$ becomes a positive real number
(see (\ref{17}) and (\ref{30})). Besides,
$\delta_\omega={3\lambda^2\over 8\pi G}$ implies that $A_\omega$
is positive. Therefore the stability conditions are satisfied.
Note that in \cite{stab2} it was proposed that $f(R)$ models with
$B<0$, where $B$ is defined through
\begin{equation}\label{64}
B\equiv {{d^2f\over d R^2}\over {df\over dR}}{dR\over d \ln
a}\left({d\ln H\over d\ln a}\right)^{-1},
\end{equation}
are unstable to linear perturbations at high curvature, this leads
us to take $\lambda>0.5$ which is in in agreement with
$0.5<\lambda<0.64$ proposed above.

Following (\ref{9}) and (\ref{16}), the energy densities can be
obtained as
\begin{equation}\label{29}
\rho_i=\delta_i A_it^{-3\lambda(1+\gamma_i)}.
\end{equation}
For radiation component this yields
\begin{equation}\label{30}
\rho_\mathcal{R}={3\over 8\pi G}(-10\lambda^2+8\lambda-1)
{\lambda^{2\lambda}}\left(6(2\lambda-1)\right)^{2\lambda-1}A_\mathcal{R}t^{-4\lambda}.
\end{equation}
On the other hand the temperature of the radiation component is
given \cite{kolb}
\begin{equation}\label{31}
\rho_\mathcal{R}=\epsilon_\mathcal{R} T^4,
\end{equation}
where $\epsilon_\mathcal{R}={\pi^2 \over 30}g_\star$ and $g_\star$
is the total degrees of freedom of effective massless particles
contributing in radiation component. Hence, in (\ref{21}),
$\theta_\mathcal{R}=4$. By using (\ref{31}) and (\ref{29}) we find
out
\begin{equation}\label{32}
T=\left({\delta_\mathcal{R}A_\mathcal{R}\over
\epsilon_\mathcal{R}}\right)^{1\over 4}t^{-{n_\mathcal{R}\over
2}}.
\end{equation}
By substituting (\ref{32}) into (\ref{29}) one gets
\begin{equation}\label{33}
\rho_\omega= \delta_\omega
A_\omega\left({\epsilon_\mathcal{R}\over
\delta_\mathcal{R}A_\mathcal{R}}\right)^ {n_\omega\over
n_\mathcal{R}}T^{4{n_\omega\over n_\mathcal{R}}}
\end{equation}
By considering the assumption (\ref{21}) we find
\begin{equation}\label{34}
\epsilon_\omega=\delta_\omega
A_\omega\left({\epsilon_\mathcal{R}\over
\delta_\mathcal{R}A_\mathcal{R}}\right)^ {n_\omega\over
n_\mathcal{R}},\,\,\, \theta_\omega=4{n_\omega\over
n_\mathcal{R}}.
\end{equation}

If at a time denoted by $t=t_{RD}$, the energy density of
radiation component becomes equal to the other component:
\begin{equation}\label{35}
\rho_\omega(t_{RD})=\rho_\mathcal{R}(t_{RD})=\epsilon_\mathcal{R}T_{RD}^4,
\end{equation}
where $T_{RD}$ is the radiation temperature at $t_{RD}$, then the
solution of (\ref{25}) may be written as
\begin{eqnarray}\label{36}
\rho_\omega=\epsilon_\mathcal{R} T_{RD}^4\left({T\over
T_{RD}}\right)^{4{n_\omega\over n_\mathcal{R}}}.
\end{eqnarray}
The temperature $T_{RD}$ can be determined in terms of the
parameters of the model:
\begin{equation}\label{37}
T_{RD}=\epsilon_\mathcal{R}^{-{1\over 4}}\left(\delta_\omega
A_\omega\right)^{n_\mathcal{R}\over
4(n_\mathcal{R}-n_\omega)}\left(\delta_\mathcal{R}
A_\mathcal{R}\right)^{n_\omega\over 4(n_\omega-n_\mathcal{R})}.
\end{equation}

As an application, we can use these results to study a simple
gravitational baryogenesis model \cite{david}, in the context of
modified theory of gravity in the early universe composed
approximately of radiation and a nonthermal component described by
(\ref{25}). The key ingredient in this theory is the coupling of
derivative of the Ricci scalar curvature and the baryon number
current
\begin{equation}\label{38}
{\varepsilon\over \Lambda^2}\int d^4x\sqrt{-g}(\partial_\mu
R)J^\mu,
\end{equation}
where $\Lambda$ is a cutoff characterizing the scale of the energy
in the effective theory and $\varepsilon=\pm 1$. (\ref{38})
dynamically violates CPT giving rise to the baryon asymmetry. In a
universe with spatially constant $R$, to obtain the chemical
potential for baryon ($\mu_B$) and antibaryons ($\mu_{\bar{B}}$),
we use
\begin{equation}\label{39}
{1\over \Lambda^2}(\partial_\mu R)J^\mu={1\over
\Lambda^2}\dot{R}(n_B-n_{\bar B}),
\end{equation}
where $n_B$ and $n_{\bar{B}}$ are the baryon and antibaryon number
densities respectively. Therefore there is an energy shift,
${2\varepsilon\dot{R}\over\Lambda^2}$, for a baryon with respect
to an antibaryon. We can assign a chemical potential to baryons:
$\mu_B=-\mu_{\bar{B}}=-{\varepsilon\dot{R}\over \Lambda^2}$. So,
in thermal equilibrium there will be a nonzero baryon number
density given by:
\begin{equation}\label{40}
n_b=n_B-n_{\bar{B}}={g_bT^3\over 6\pi^2}\left(\pi^2{\mu_B\over
T}+({\mu_B\over T})^3\right),
\end{equation}
where $g_b\sim \mathcal{O}(1)$ is the number of internal degrees
of freedom of baryons. The entropy density of the universe is
given by $s={2\pi^2\over 45} g_sT^3$, where $g_s\simeq 106$
indicates the total degrees of freedom for relativistic particles
contributing to the entropy of the universe \cite{kolb}. In the
expanding universe the baryon number violation decouples at a
temperature denoted by $T_D$ and a net baryon asymmetry remains.
The ratio ${n_b\over s}$ in the limit $T_D\gg m_b$ ($m_b$
indicates the baryon mass), and $T_D\gg \mu_b$ is then:
\begin{equation}\label{41}
{n_b\over s}\simeq -\varepsilon{15g_b\over 4\pi^2g_s}{\dot{R}\over
\Lambda^2T}|_{T_D}.
\end{equation}
Introduction of $\varepsilon$ gives us the possibility to choose
the appropriate sign for $n_b$. In our model $\dot{R}$ is
determined as
\begin{equation}\label{42}
\dot{R}=-6n_\mathcal{R}(1-n_\mathcal{R})\left(\epsilon_\mathcal{R}\over
{\delta_\mathcal{R} A_\mathcal{R}}\right)^{3\over
2n_\mathcal{R}}T_D^{6\over n_\mathcal{R}}.
\end{equation}
To derive (\ref{42}) we have used the fact that the decoupling
time is
\begin{equation}\label{43}
t_{D}=\left(\epsilon_\mathcal{R}\over {\delta_\mathcal{R}
A_\mathcal{R}}\right)^{-{1\over 2n_\mathcal{R}}}T_D^{-{2\over
n_\mathcal{R}}}.
\end{equation}
By putting (\ref{42}) into (\ref{41}) we can determine the baryon
asymmetry
\begin{equation}\label{44}
{n_B\over s}\sim \varepsilon{0.02\over
\Lambda^2}n_\mathcal{R}(1-n_\mathcal{R})\left(\epsilon_\mathcal{R}\over
{\delta_\mathcal{R} A_\mathcal{R}}\right)^{3\over
2n_\mathcal{R}}T_D^{6-n_\mathcal{R}\over n_\mathcal{R}}.
\end{equation}
Note that (\ref{44}) is valid for radiation dominated epoch as
well as for non-radiation dominated era provided that the EOS
parameter of non radiation component satisfies (\ref{28}).

\end{document}